\documentclass[prl,superscriptaddress,showpacs,twocolumn]{revtex4}

\usepackage{graphicx}

\usepackage{amssymb,amsmath}
\usepackage[raggedright,hang,nooneline]{subfigure}
\usepackage[dvips]{color}

\newcommand{\fref}[1]{Fig.~\ref{#1}}

\newcommand{\im}{%
           \imath}
\newcommand{\bra}[1]{\ensuremath{\langle #1|}}
\newcommand{\ket}[1]{\ensuremath{|#1\rangle}}

\newcommand{\fermi}[1]{%
        \hbox{f($#1$)}}
\newcommand{\dif}{%
        \hbox{d}}           
\newcommand{\tr}{%
        \hbox{ tr}}

\def\eg{{e.g.\ }}

\def\t2g{\ensuremath{t_{2g}}}
\def\a1g{\ensuremath{a_{1g}}}

\newcommand{\svek}{%
        \mathbf}
\newcommand{\vek}[1]{%
        \hbox{\textbf #1}}

\newcommand{\op}[1]{%
        \hbox{\textbf #1}}
\newcommand{\pr}{%
        ^\prime}

\def\vo2{VO\ensuremath{_2}}
\def\tio2{TiO\ensuremath{_2}}
\def\sio2{SiO\ensuremath{_2}}
\def\VO2{VO\ensuremath{_2}}
\def\v2o3{V$_2$O$_3$}
\def\V2O3{V$_2$O$_3$}
\def\vcro2{V$_{1-x}$Cr$_x$O$_2$}
\def\etal{{\it et~al.}}

\def\dvo2{\ensuremath{d_{\hbox{\tiny VO}_2}}}
\def\dtio2{\ensuremath{d_{\hbox{\tiny TiO}_2}}}
\def\dsio2{\ensuremath{d_{\hbox{\tiny SiO}_2}}}

\def\XXint#1#2#3{{\setbox0=\hbox{$#1{#2#3}{\int}$}
\vcenter{\hbox{$#2#3$}}\kern-.5\wd0}}

\begin{document}

\title{Materials Design using Correlated Oxides:
Optical Properties of Vanadium Dioxide
}

\author{Jan M. Tomczak}
\affiliation{Research Institute for Computational Sciences, AIST, Tsukuba, 305-8568 Japan}
\affiliation{Japan Science and Technology Agency, CREST}
\author{Silke Biermann}
\affiliation{Centre de Physique Th{\'e}orique, Ecole Polytechnique, CNRS,
91128 Palaiseau Cedex, France}
\affiliation{Japan Science and Technology Agency, CREST}

\begin{abstract}
Materials with strong electronic Coulomb interactions
play an increasing role in modern materials applications. 
``Thermochromic'' 
systems, 
which exhibit thermally induced changes in their optical response, provide a particularly interesting case.
The optical switching associated with the
metal-insulator transition of vanadium dioxide (\vo2), for example,
has been proposed for use in 
``intelligent'' windows,
which selectively filter 
radiative heat in hot weather conditions.
In this work, we develop the theoretical tools for
describing such 
a behavior.
Using a novel scheme for the calculation of the optical
conductivity of correlated materials, we obtain 
quantitative agreement with experiments
for both phases of \vo2.
On the example of an optimized energy-saving window setup, 
we further demonstrate that theoretical materials design  has now come 
into reach, even for the particularly challenging class
of correlated electron systems.
\end{abstract}

\maketitle

The concerted behavior of electrons in 
materials
with strong electronic Coulomb interactions
causes an extreme sensitivity
to external stimuli, such as
temperature, pressure or external fields. 
Heating insulating SmNiO$_3$ beyond 400 K or 
applying a pressure of just a few kbar to the Mott insulator
(V$_{1-x}$Cr$_x$)$_2$O$_3$ (x=0.01), e.g.,
make the materials undergo transitions to metallic states~\cite{imada}.
This tuneability of 
fundamental properties 
is both a harbinger for diverse technological
applications and a challenge for a theoretical description.
Indeed, many materials
used in modern applications fall in 
this class of 
strongly correlated materials where electron-electron interactions
profoundly modify (if not invalidate) a pure band picture.
State-of-the-art first 
principles methods, such as density functional theory, are then no longer sufficient to predict
physical properties of these compounds.

An increasing role  
is nowadays played by
artificial
structures, ranging from spintronic multi-layers~\cite{PhysRevLett.61.2472,PhysRevB.39.4828}
to functional surfaces, where appropriate coatings e.g.\ provide a self-cleaning
mechanism~\cite{selfcleaning}.
However, the huge freedom in the design,  
which concerns
not only  
the chemical composition, the doping and the growth conditions, but
also  
the geometry of the setup (e.g.\ the layer thicknesses),
makes the search for devices with specific electronic properties
a tedious task.
Ideally,  the quest for promising materials and setups 
should therefore be seconded by theoretical predictions.
Yet, the abundance of fruitful experimental work 
on tailoring e.g.\ materials with specific optical 
absorption profiles 
is met with only scarce contributions from
theoretical investigations~\cite{transcond}.

Here, we demonstrate that even for the particularly challenging class
of correlated materials a {\it quantitative} description, and thus
predictions, are possible. 
We introduce a novel
scheme for the calculation of optical properties, 
which can
in principle be employed in conjunction with any electronic structure
technique that uses localized basis functions. 
We shall demonstrate the power of the approach within the framework
of the combined density functional - dynamical mean field method
``LDA+DMFT''~\cite{vollkot}, on the example of the optical
conductivity of vanadium dioxide, \vo2. We further investigate \vo2
based heterostructures and show that theoretical materials design 
of correlated 
material-derived devices is coming into reach.

Within 
dynamical mean field theory, 
emphasis is commonly put on {\it spectral} properties, and
 the evaluation of 
observables other than spectral functions is a rather new advancement in the realistic context.
Yet, as explained above, it is the {\it response} behavior of correlated systems that is promising for applications.
In this vein, recent pioneering work~\cite{bluemer,oudovenko:125112,haule:036401,1367-2630-7-1-188,Baldassarre_v2o3}
 has
successfully described
 qualitative features of the optical response of several correlated materials.
The calculation of 
absolute values in theoretical response functions -- a prerequisite for quantitative theoretical materials design --
 has, however, turned out to be a formidable challenge. This is mainly
due to the sensitivity of the obtained results on the accuracy
of the transition matrix elements~\cite{me_phd}.

\begin{figure*}[t]
  \begin{center}
    \mbox{
\subfigure[$\qquad$]{\scalebox{0.35}{\includegraphics[angle=-90,width=\textwidth]{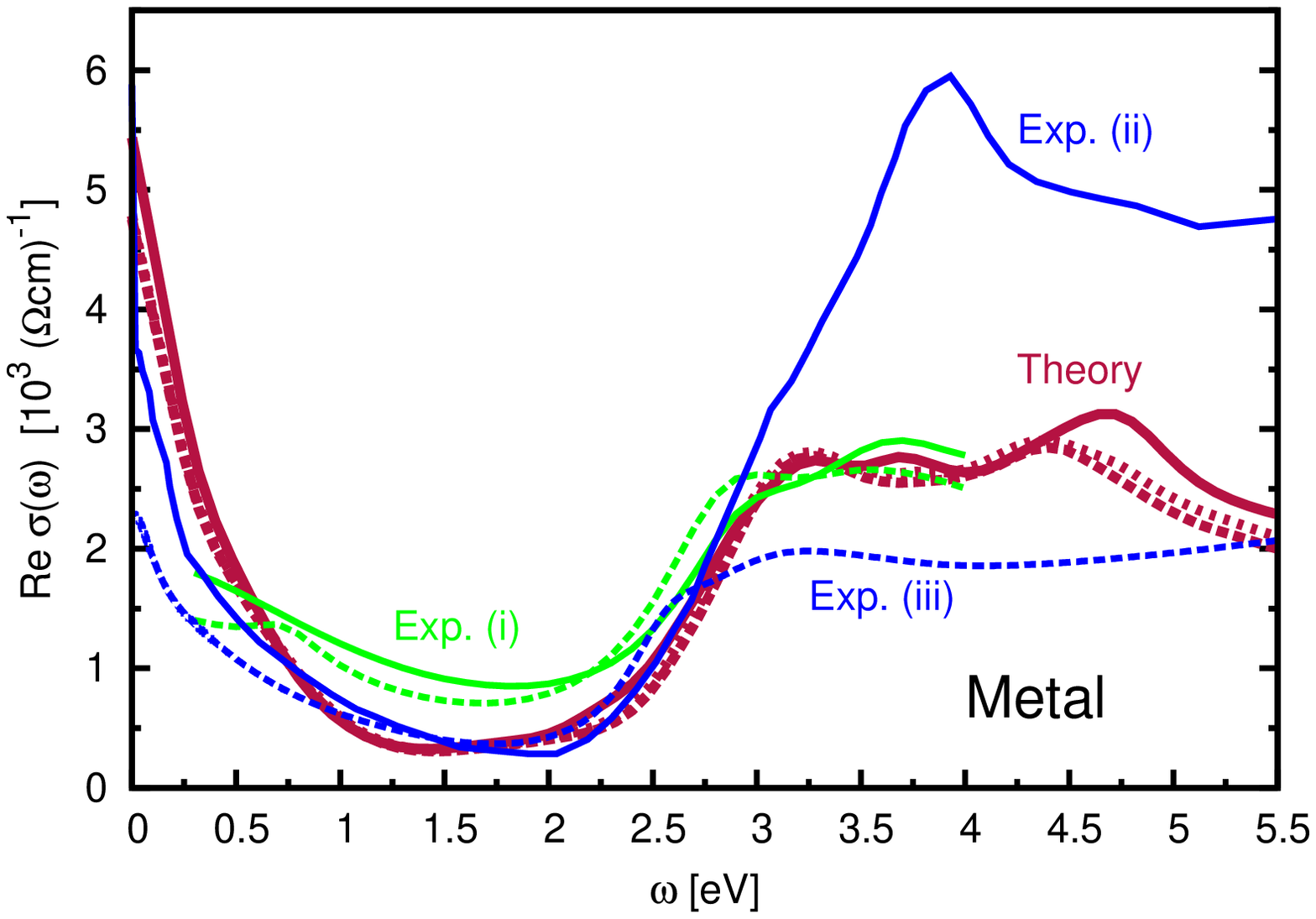}}} 
\hspace{1.25cm}
\subfigure[$\qquad$]{\scalebox{0.35}{\includegraphics[angle=-90,width=\textwidth]{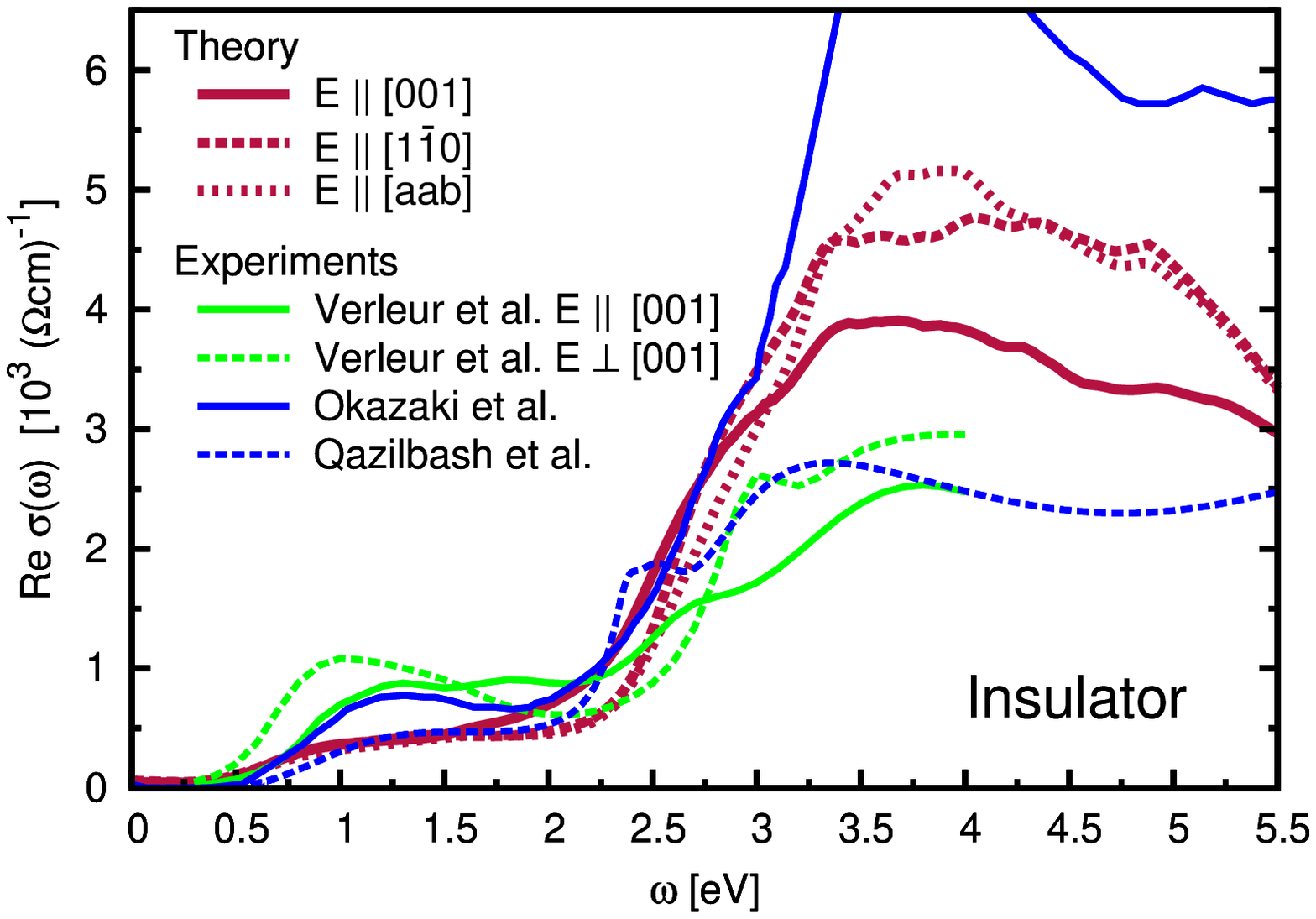}}} 
      }
    \caption{Optical conductivity of (a) metallic, (b) insulating \vo2\ for polarizations $E$. Theory (red)
    ([aab]=[0.85 0.85 0.53]), experimental data (i)
      single crystals~\cite{PhysRev.172.788}
       (green), (ii) thin film~\cite{PhysRevB.73.165116}
       (solid blue), (iii)
      polycrystalline film~\cite{qazilbash:205118}
       (dashed blue).
}
    \label{fig1}
  \end{center}
\end{figure*}
In the framework of 
linear response theory, and when neglecting vertex-corrections~\cite{PhysRevLett.64.1990}, 
the optical conductivity can be expressed as (for reviews see~\cite{millis_review,me_phd})
\begin{eqnarray}\label{oc}
Re\,\sigma ^{\alpha\beta} (\omega)&=&\frac{2\pi e ^2\hbar}{V}\sum_{\svek{k}}\int\dif\omega\pr\;\frac{\fermi{\omega\pr}-\fermi{\omega\pr+\omega}}{\omega} \nonumber\\
&\times&
\tr\biggl\{ A_{\svek{k}}(\omega\pr+\omega) v_{\svek{k},\alpha} A_{\svek{k}}(\omega\pr)  v_{\svek{k},\beta} \biggr\}\label{oc_para2}
\end{eqnarray}
Correlation effects enter the calculation via the spectral functions
$A_{\svek{k}}(\omega)$, while
the Fermi velocities $v_{\svek{k}, \alpha}=\frac{1}{m}\bra{\svek{k}L\pr}\mathcal{P}_\alpha\ket{\svek{k}L}$, matrix elements of the momentum operator 
$\mathcal{P}$, which weight the different transitions,
are determined by the band-structure.
Both, spectral functions and velocities are matrices in the basis
of localized orbitals, indexed 
by $L=(n,l,m,\gamma)$, with the
usual quantum numbers $(n,l,m)$, while $\gamma$ denotes the 
atoms in the unit cell~:
$\ket{\svek{k}L}$ is the 
Fourier transform of the Wannier function
$\chi_{\svek{R}L}^{\phantom{b}}(\vek{r})$ localized at 
atom $\gamma$ in the unit cell $\vek{R}$.
 The Fermi functions $\fermi{\omega}$ select the range of occupied and empty energies, respectively,
$V$ is the unit-cell volume, 
$\alpha$,$\beta$ denote cartesian coordinates, and $Re\,\sigma ^{\alpha\beta}$ is the response in $\alpha-$direction for a light polarization $E$ along $\beta$.
Approaches such as LDA+DMFT~\cite{vollkot}
 require the use of localized, Wannier-like, basis sets, rendering the evaluation of the full matrix element 
tedious.
To this end, we developed a ``generalized Peierls'' approach that extends the formalism for
 lattice models (see the review~\cite{millis_review})
  to the realistic
 case of multi-atomic unit cells.
We find that
\begin{eqnarray}
v_{\svek{k},\alpha}^{L\pr L} 
&=&
{\frac{1}{\hbar} \biggl(        \partial_{k_\alpha}\op{H}^{L\pr
                   L}_{\svek{k}} -\im
                   (\rho_{L\pr}^\alpha-\rho_L^\alpha)\op{H}^{L\pr
                   L}_{\svek{k}}        
 \biggr)} + {\mathcal{F}_{\op{H}}\left[\{\chi_{\svek{R}L}^{\phantom{b}} \}\right]}\nonumber\\
\end{eqnarray}
Here, $\rho_L$ denotes the position of an atom within the unit-cell.
The term in brackets, 
which is used in the actual calculations, 
we refer to as the ``generalized Peierls'' term~:
While the derivative term is the common Fermi velocity, the one proportional to the Hamiltonian originates from the generalization to realistic multi-atomic unit-cells. The correction term that recovers the full matrix element is denoted 
$\mathcal{F}$ (for its explicit form see~\cite{me_phd}).
The 
latter reduces to purely atomic transitions, $(\svek{R},\gamma)=(\svek{R}\pr,\gamma\pr)$,
in the limit of strongly localized orbitals. In other words, the accuracy of the approach is controlled by the choice of basis functions.
This generalized Peierls approach 
is thus expected to be a 
good approximation 
 for systems with localized orbitals, such as the 3d or 4f orbitals
in transition metal or lanthanide/actinide compounds. This
expectation is indeed true as can be verified numerically~\cite{me_phd}.
From the optical conductivity, one can calculate the
specular reflectivity and transmittance.
Therewith, also the specular color of 
the material becomes accessible~\footnote{We
employ the CIE 1964 conventions, along with the daylight illuminant D65, see \eg
K. Nassau.
{\em The Physics and Chemistry of Color: The Fifteen Causes of Color}. 
(Wiley Series in Pure \& Applied Optics, 2001).
}.

Our focus compound, \vo2, has attracted a lot of attention recently~\cite{PeterBaum11022007,Qazilbash12142007}.
It
is among those materials in which correlation
effects play a decisive role, to the extent that standard band-structure
approaches fail to capture even most basic experimental facts~:
In metallic \vo2, important incoherent spectral features witnessed by
photoemission experiments are 
absent in band theory, and the metal-insulator transition at T$_c$=340~K
is missed altogether. For reviews see e.g.~\cite{eyert_vo2,me_phd} 
and references therein.
Hence, any description of the thermochromic properties of this material must foot on an electronic structure
approach that goes beyond band-theory and masters the many-body effects at work.
Indeed, LDA+DMFT 
results for the spectral properties of \vo2
agree well with experimental findings in
both, the metallic~\cite{0295-5075-69-6-984,liebsch:085109, biermann:026404} 
 and the insulating phase~\cite{biermann:026404}.
Footing on the electronic structure calculation of Ref.~\cite{biermann:026404}
and our recent extension thereof~\cite{tomczak_vo2_proc,me_vo2},
we first employ our 
scheme to access optical properties of bulk \vo2\ both above and below its metal-insulator transition. 

\begin{figure*}[t!h]
\begin{minipage}[c]{0.32\textwidth}
\subfigure[$\qquad$]{\includegraphics[angle=-90,width=.85\textwidth]{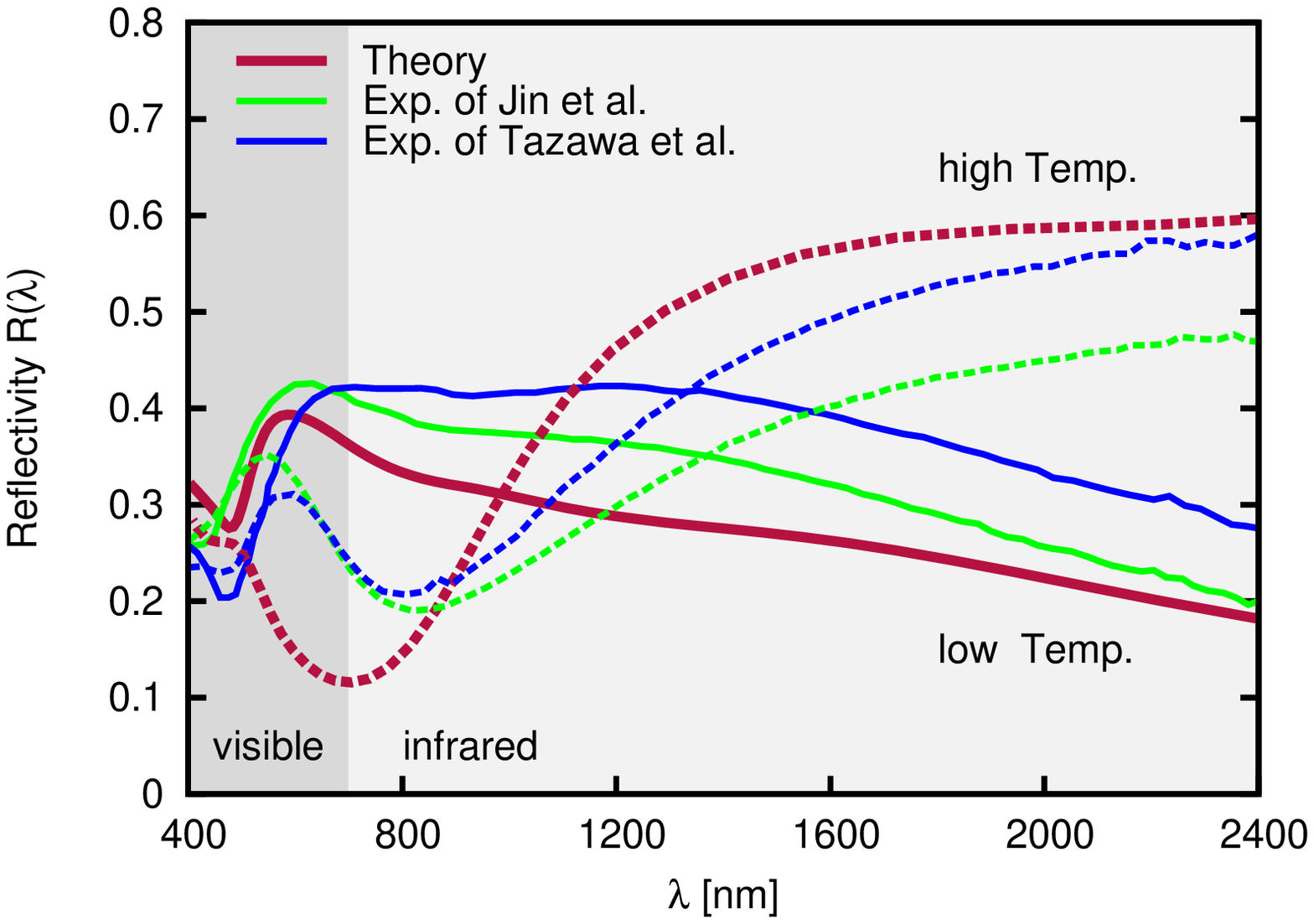}}
\end{minipage}
\begin{minipage}[c]{0.67\textwidth}
\subfigure[$\qquad$]{
\includegraphics[angle=-90,width=.45\textwidth]{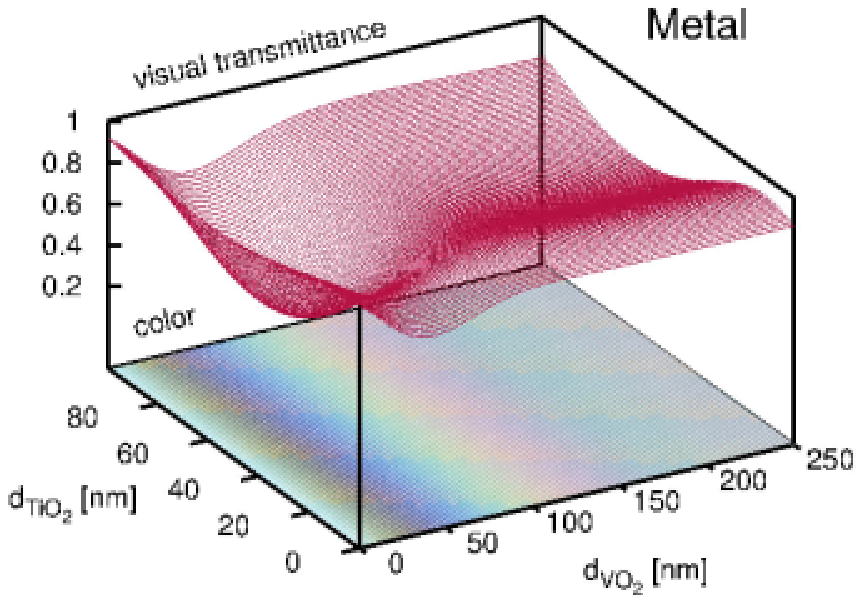}
}
\hspace{-0.75cm}
\subfigure[$\qquad$]{
\includegraphics[angle=-90,width=.45\textwidth]{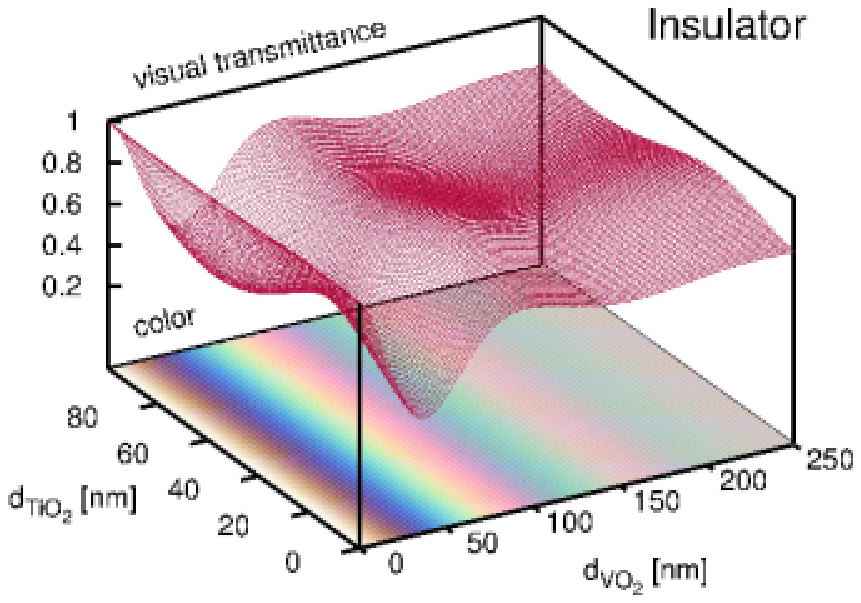}
}
\end{minipage}
    \caption[bla]{Setups of an intelligent window.
    (a) \vo2 on \sio2. Reflectivity
      at high (dashed) and low temperatures (solid).
      Theory (red) : 60 nm \vo2\ on
      \sio2. Experiments :
      50 nm \vo2\ on SiO$_2$~\cite{jinvo2}
       (green), 50 nm \vo2\ on Pyrex glass~\cite{Tazawa:98}  (blue).
       (b), (c) Setup with antireflexion coating : \tio2 / \vo2 on \sio2.       
   Shown is the normalized visible transmittance (see text)
   an the corresponding color of the transmitted light   
   for  (b) high and (c) low temperature 
   as a function of the layer thicknesses \dvo2\ and \dtio2. 
All theoretical data for $E \parallel [1\bar{1}0]$ polarization, and a 3mm \sio2\ substrate.}
    \label{fig2}
  \end{figure*}

\fref{fig1}(a) shows our results for the optical conductivity of 
metallic \vo2
 as a function of frequency and in comparison with 
experimental data~\cite{PhysRev.172.788,qazilbash:205118,PhysRevB.73.165116},
 see also~\cite{Qazilbash12142007}.
As can be expected from the crystal structure~\cite{eyert_vo2}, 
the optical response
depends only weakly on the light polarization. 
The Drude-like metallic response is caused by transitions between narrow vanadium 3d
excitations near the Fermi level, and
 thus affects only the low
infra-red regime -- a crucial observation 
as seen in the following. The shoulder at 1.75~eV 
still stems from intra-vanadium 3d contributions, while transitions involving 
oxygen 2p orbitals set in at 2~eV, henceforth constituting
the major spectral weight up to the highest energies of the calculation.
Also for insulating \vo2, see \fref{fig1}(b), the results are
in good agreement with 
experiments. This
time, a slight polarization dependence is seen in both, experiment and
theory,
owing to the change in 
crystal symmetry. Indeed,  vanadium
atoms pair up in the insulator to form dimers along the c-axis,
leading to the formation of bonding/anti-bonding states for 3d
orbitals~\cite{goodenough_vo2}. 
Optical transitions between these orbitals result,
in the corresponding energy range ($\omega$=1.5-2.5~eV),
in 
a higher amplitude of the conductivity 
for a light polarization parallel
 to the c-axis than for other directions~\cite{me_phd}. 

Having established the theoretical optical response of
{\it bulk} \vo2, and thus verified that our scheme can {\it quantitatively} predict optical properties of correlated materials, we now investigate the
possibilities of \vo2-based intelligent
window coatings~\cite{Babulanam:1987,windowvo2}. 
The effect to be exploited can already be seen
 in the above bulk responses~:
The conductivities of both phases (\fref{fig1}(a), (b))
exhibit a close similarity in the
range of visible light ($\omega$=1.7-3.0~eV), whereas
in the infra-red regime ($\omega$$<$1.7~eV) 
 a pronounced switching occurs across the 
  transition.
As a result, heat radiation 
can pass at low external
temperatures, while 
its transmission is hindered
above T$_c$. 
The insensitivity to temperature for visible light, in conjunction with the selectivity of the response to infrared radiation, is an
 essential feature of an intelligent window setup.
Yet, for an applicable realization, other important
 requirements have to be met.
First of all, the switching of the window 
 has to occur at a relevant,
that is  ambient, temperature.
Also, the total transmittance of \vo2-films in the visible range
needs improvement~\cite{Babulanam:1987}, 
and the transmitted visible light should 
be uniform 
in order to provide a
colorless vision.
Experimentalists have addressed these issues and have proposed potential solutions~\cite{transcond}~:
Diverse dopings, $M_xV_{1-x}O_2$, were proved to influence T$_c$, with Tungsten ($M$=$W$) being the most efficient~: A doping of only 6\%
results in T$_c$$\approx$ 20$^\circ$C~\cite{Sobhan}.
 However, this causes a
deterioration of the infrared switching. Fluorine doping, on the other hand, improves on the switching properties, while also reducing T$_c$~\cite{tfcd,tfcd2}. 
An increase in the overall visible transmittance can also be achieved
without modifying the intrinsic properties of the material itself,
but 
by adding
antireflexion coatings,
for example 
TiO$_2$~\cite{jinvo2}.

Here, we address the optical properties of window coatings from the
theoretical perspective.
In doing so, we assume that the specular response of \vo2\ layers
is 
well-described by the optical properties of the bulk, and we use geometrical optics to deduce the properties of layered structures. 
First, we consider the most simple of all setups, which consists of a single \vo2-layer (of thickness \dvo2) on a glass substrate
~\footnote{We suppose
   quartz glass, \sio2. All auxiliary refractive indices 
   are taken from {\em Handbook of Optical Constants of Solids} by Edward~D. Palik,
Academic Press, 1985.}.
Such a window has been experimentally investigated by Tazawa \etal~\cite{Tazawa:98}
and Jin \etal~\cite{jinvo2}.
 In \fref{fig2}a
  we show their measured reflectivity data 
 as a function of wavelength, in comparison with our theoretical results~:
At low temperatures 
(insulating \vo2), the calculated reflectivity
is in quantitative agreement with the experimental data.
In the visible range ($\lambda$$\approx$400-700~nm), the reflectivity strongly depends  on the wavelength.
Therefore, the current window
will filter certain wavelengths more than others, resulting in an illumination of a certain color
-- an obvious drawback.
 Moreover, the reflectivity in this region is rather elevated, causing
poor global transmission.
In the infrared regime ($\lambda$$>$700~nm) -- and beyond -- the reflectivity decreases, and
radiation that causes greenhouse heating can pass the setup.
At high temperatures, 
the infrared reflectivity 
switches to a rather elevated value, thus filtering
heat radiation.
The changes in the visible region are less
pronounced, but still perceptible, and both, the degree
of transparency and the color  
change
 through the transition.
The current setup is thus not yet
 suited for applications.

We therefore now investigate a more complicated setup~: 
An additional (rutile) \tio2-coating
is added on 
top of the \vo2-layer, with the objective of serving as an antireflexion filter~\cite{jinvo2}.  
With the thicknesses \dvo2\ and \dtio2, the geometry of the current setup 
has two parameters that can be used to optimize the desired optical properties.
Since, however, each variation of them requires the production of a new individual sample 
under comparable deposition conditions, along with a careful structural
characterization in order to guarantee that differences in the optical behavior are genuine
and not related to variations of the sample quality,
the experimental expenditure is tremendous.
This led Jin \etal~\cite{jinvo2}
 to first estimate a highly transmitting setup by using tabulated
refractive indices and to produce and measure only {\it one} such sample.
Here, we shall use our theoretical results on \vo2\ to not only optimize the geometry (\dtio2, \dvo2)
with respect to the total visible transmittance, but 
to also investigate the transmission color.
\fref{fig2}
 displays the normalized visible specular transmittance
\footnote{$\int_{400nm}^{700nm} d\lambda
S(\lambda)T(\lambda)/\int_{400nm}^{700nm} d\lambda S(\lambda)$, with the 
spectrum of the light
source $S(\lambda)$, and 
the transmittance $T=1-R$, $R$ being the
  specular reflexion given by Fresnel's formulae.
We thus  neglect absorption due to inhomogeneities that lead to diffuse
  reflexion. This is justified for our applications to 
 windows. Also, \vo2\ has a glossy
  appearance and hence a preponderant specular response.}
 for the window in its high (b) and low (c) temperature state, as a function of both film thicknesses. 
On the same graph 
the resulting transmission colors
are  shown.
The evolution of the light interferences in the layers 
causes 
pronounced changes in both, the overall transmittance and the color.
The coating of \vo2\ globally degrades the transparency of the bare
glass window.
An increase of the TiO$_2$-coating, on the other hand, has the potential to improve the total
transmittance. 
This can be understood from the mechanism of common 
quarter-wave filters. 
The wavelength-dependence of the 
real-part of the TiO$_2$ refractive index, $n_{\hbox{\tiny TiO}_2}(\lambda)$, results in an
  optimal quarter-wave thickness,
 $\delta_{\hbox{\tiny TiO}_2}(\lambda)$=$\lambda/(4n_{\hbox{\tiny TiO}_2}(\lambda))$,  which varies
from blue to red light only slightly from $\delta_{\hbox{\tiny TiO}_2}(\lambda)$=$40$ to $60$~nm.
This and the fact that the imaginary part of the refractive index, $k_{\hbox{\tiny TiO}_2}(\lambda)$, is negligible for visible light
explains why the color does not change significantly with \dtio2. 
While, as for TiO$_2$, the variation of the real-part of the \vo2\
refractive index yields a rather uniform ideal thickness $\delta_{\hbox{\tiny VO}_2}(\lambda)$, its
imaginary part changes significantly (by a factor of 4) within the range of visible light.
As a consequence, the color is very sensitive to \vo2-deposition. At
higher thickness \dvo2, however, this 
dependence becomes
smaller and the color lighter. 
Our theoretical transmittance profiles 
suggest relatively thick windows
to yield good visual properties. 
Indeed, at low temperatures, \fref{fig2}(c), 
the local maximum that gives the thinnest
window 
 is located at $(\dtio2 , \dvo2)$$\approx$(40~nm, 85~nm) within our calculation.
However, this setup 
is still in the regime of important color
oscillations. Given the uncertainties  in industrial deposition techniques,
it seems 
 cumbersome to consistently stabilize colorless samples.
From this point of view, a thicker \vo2-film would be desirable.
Indeed, while almost preserving the overall transmittance, a 
colorless window at low temperatures is realized 
for $(\dtio2 , \dvo2)$$\approx$(50~nm, 220~nm), or for $(\dtio2 , \dvo2)$=($\ge$$ 100~\hbox{nm},
220~\hbox{nm})$. 
In the high temperature state, \fref{fig2}(b),
 the transmittance is globally lower than at low temperatures.
Moreover, only $(\dtio2 , \dvo2)$=$(\ge 100~\hbox{nm},
220~\hbox{nm})$ yields a simultaneously high transmittance
in {\it both} states of the window.

In conclusion, we 
have presented a novel scheme for the calculation of optical
properties of correlated materials, and applied it to vanadium dioxide, \vo2.
We find the bulk optical conductivity of both phases in quantitative
agreement with experiments, and further validate our approach by
comparing the transmittance of 
a VO$_2$-layer on SiO$_2$
to experimental
data.
Finally, we optimize
the geometry of a multi-layer setup of an intelligent window, which
uses the metal-insulator transition of \vo2\ to reduce the effect of 
greenhouse heating.
This work can be considered as a proof of principle of the
feasibility of theoretical materials design, since 
our techniques can be applied to the general class of correlated materials.

We gratefully acknowledge discussions with T. Gacoin, G. Garry,
A. Georges and L. Pourovskii.
This work was supported by  
the Japan Science and Technology Agency (JST) under the CREST program,
the French ANR under project
CORRELMAT, and a computing grant by IDRIS Orsay under project number
081393.

\end{document}